\documentclass[12pt,a4paper,final]{iopart}
\usepackage{iopams}

\usepackage{graphicx}
\usepackage{dcolumn}
\usepackage{bm}
\usepackage{hyperref}
\usepackage[dvipsnames]{xcolor}
\usepackage{soul}

\hypersetup{
    colorlinks=true,
    citecolor=Blue,
    linkcolor=Blue,
    urlcolor=Blue
}
\newcommand{\mref}[2]{%
    \hyperref[{#1}]{%
        \ref*{#1}#2%
    }%
}

\begin{document}
\title{Collisions in a dual-species magneto-optical trap of molecules and atoms}

\author{S. Jurgilas$^1$, A. Chakraborty$^1$, C. J. H. Rich$^1$, B. E. Sauer$^1$, Matthew D. Frye$^2$, Jeremy M. Hutson$^2$ and M. R. Tarbutt$^1$}

\address{$^1$ Centre for Cold Matter, Blackett Laboratory, Imperial College London, Prince Consort Road, London SW7 2AZ UK}
\address{$^2$ Joint Quantum Centre (JQC) Durham-Newcastle, Department of Chemistry, Durham University, South Road, Durham DH1 3LE, UK}

\ead{m.tarbutt@imperial.ac.uk}

\begin{abstract}
We study inelastic collisions between CaF molecules and $^{87}$Rb atoms in a dual-species magneto-optical trap. The presence of atoms increases the loss rate of molecules from the trap. By measuring the loss rates and density distributions, we determine a collisional loss rate coefficient $k_{2} = (1.43 \pm 0.29) \times 10^{-10}$~cm$^{3}$/s at a temperature of 2.4~mK. We show that this is not substantially changed by light-induced collisions or by varying the populations of excited-state atoms and molecules. The observed loss rate is close to the universal rate expected in the presence of fast loss at short range, and can be explained by rotation-changing collisions in the ground electronic state.
\end{abstract}


\section{Introduction}

The study of collisions is an important topic in the field of cold atoms and molecules. Collisions are essential for sympathetic cooling, evaporative cooling and the association of atoms to form molecules. Low-temperature collisions are sensitive to the long-range  part of the interaction potential and can be used to determine the energies of the highest-lying bound states of the potential. Inelastic and reactive collisions often limit the densities and lifetimes that can be reached with trapped atoms and molecules. Understanding such collisions is important and may suggest methods of controlling the collisions and avoiding the limitations that they otherwise impose. 

At temperatures in the range 10~mK to 1~K, collisions have been studied using crossed and merged beams~\cite{Henson2012, Klein2016, Wu2017, Jankunas2015, deJongh2020}, and using mixtures loaded into traps by buffer-gas cooling or deceleration~\cite{Hummon2011, Parazolli2011, Fitch2020, Segev2019, Reens2017}. At lower temperatures, collisions between laser-cooled atoms have been studied extensively, both by theory and experiment~\cite{Weiner1999}. More recently, collisions between ultracold molecules~\cite{Krems2009} have received considerable attention. These collisions result in rapid loss from optical traps, due either to chemical reactions or the formation of long-lived complexes that are excited by the trap laser~\cite{Ospelkaus2010, Christianen2019, Gregory2020, Anderegg2019, Cheuk2020}. Recently, elastic collisions between ultracold atoms and molecules in optical traps have been observed and used to bring the species into thermal equilibrium~\cite{Son2020, Tobias2020}, and inelastic collisions between laser-cooled atoms and molecules have been studied in a magnetic trap~\cite{Jurgilas2021}. 

\begin{table}[b]
\centering
\begin{tabular}{|c|c|c|}
\hline
Species & Structure & Frequency   \\
\hline
CaF & Vibration & 17.5~THz \\
CaF & Rotation & 20.5~GHz \\
CaF & Hyperfine & 148~MHz \\
CaF* & Lambda doubling & 1.36~GHz \\
Rb & Hyperfine & 6.83~GHz \\
Rb* & Fine & 7.12~THz \\
Rb* & Hyperfine & 496~MHz \\

\hline
\end{tabular}
\caption{Some relevant energies, expressed as frequencies. An asterisk denotes an electronic excited state.}
\label{tab:energies}
\end{table}

In this paper, we present the first study of collisions between atoms and molecules in a magneto-optical trap (MOT). We create a dual-species MOT of $^{87}$Rb and CaF and study the loss of molecules from the trap due to collisions with the atoms. We measure the loss rate coefficient, investigate which of the many possible processes are dominant, and interpret our findings in the context of simple theoretical models. The investigation of atom-atom collisions in MOTs has a long history~\cite{Prentiss1988, Sesko1989, Wallace1992, Kawanaka1993, Marcassa1993, Ritchie1995, Gensemer1997}. In these atomic MOTs, collisional loss rate coefficients are found to vary over several orders of magnitude and often show extremely strong dependence on the MOT intensity and detuning. In some cases, rate coefficients exceeding $10^{-10}$~cm$^{3}$~s$^{-1}$ have been measured. At low laser intensities, where the excited-state fraction is small and the trap depth is low, the dominant collisional loss mechanism tends to be hyperfine-changing collisions between ground-state atoms that release enough energy to eject the atoms from the MOT. At higher intensities, processes where a ground-state atom collides with an excited-state atom become important. Fine-structure-changing collisions can occur, releasing the energy of the excited-state fine-structure interval, which is typically much larger than the trap depth. In addition, a colliding pair may absorb a photon at intermediate range; the resulting molecule may spontaneously emit a photon of lower energy at short range, with the energy difference converted into kinetic energy greater than the trap depth. This is known as radiative escape.  

All the processes discussed above can also happen for the atom-molecule mixtures we study, but there can also be other processes. In particular, there can be rotation-changing collisions since laser cooling proceeds from the first rotationally excited state, and vibration-changing collisions since the MOT contains a mixture of molecules in a few vibrational states. In addition, a collision with an excited-state molecule can transfer the molecule to the opposite-parity component of the $\Lambda$ doublet, which will eject it from the MOT. Molecules might also be lost due to light-induced processes analogous to radiative escape. Table \ref{tab:energies} summarizes the relevant energy scales. These may be compared to the trap depth of the CaF MOT, which in the current experiments is $U_{\rm trap}/h \approx 1.6$~GHz. In making this comparison, one should note that in a CaF-Rb collision that releases an energy $E_{\rm int}$ much larger than the initial kinetic energies, the CaF molecule gains a kinetic energy of $0.6 E_{\rm int}$.

\section{Method}

\begin{figure}[tb]
\centering
    \includegraphics[width=0.85\textwidth]{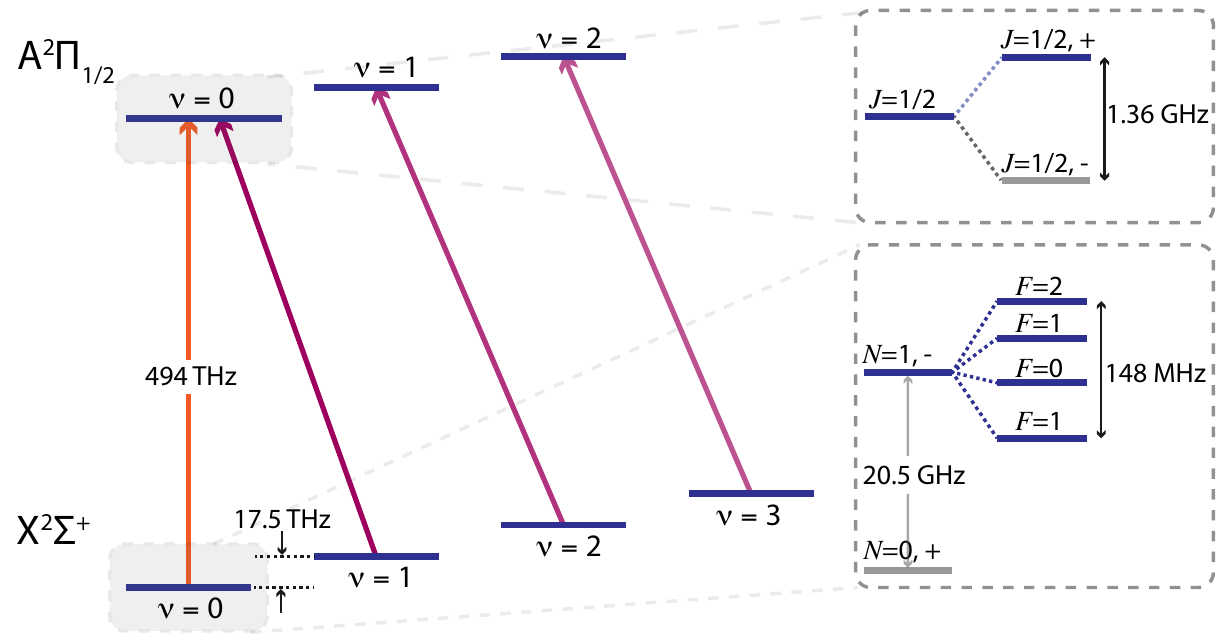}
    \caption{Relevant energy levels of CaF. Levels shown in blue are the ones used for laser cooling and magneto-optical trapping. Coloured arrows show the laser transitions driven. Collision-induced transitions to the levels shown in grey will remove molecules from the MOT. Vibrational levels are labelled by $v$. In the $X^{2}\Sigma^{+}$ state, levels are labelled as $N,p$ where $N$ is the rotational quantum number and $p$ is the parity. In the $A^{2}\Pi_{1/2}$ state, levels are labelled by $J$, the total angular momentum excluding nuclear spin. Each is split into two states of opposite parity ($\Lambda$ doubling). Hyperfine components are labelled by the total angular momentum quantum number $F$. The level structure of the other vibrational states is similar to the ones shown for $v=0$.
    \label{fig:levels}}
\end{figure}

Figure \ref{fig:levels} illustrates the relevant energy levels of CaF. The main cooling light addresses the $A^{2}\Pi_{1/2}(v'=0,J'=1/2,+)\leftarrow X^{2}\Sigma^+(v=0,N=1)$ transition at 606~nm. At the molecules, the total intensity of this light is $I_{00}$. It is combined with light from three vibrational repump lasers that maintain the molecules within the cooling cycle.  Radiofrequency sidebands are added to all four lasers to address all hyperfine components of the transitions. Because both $v=0$ and $v=1$ levels of $X^{2}\Sigma^+$ are coupled to the same excited state using roughly equal laser intensities, the populations in these states are approximately equal. There is very little population in any of the other vibrational states. Within $X^{2}\Sigma^+$ the only populated rotational state is $N=1$. As can be seen in the figure, the ground state of the laser-cooling scheme is not the lowest rotational level, and the excited state is the $\Lambda$-doublet component of higher energy. Consequently, collisions that change the rotational state or ones that change the $\Lambda$-doublet component can both occur.

\begin{figure}[tb]
\centering
    \includegraphics[width=0.7\textwidth]{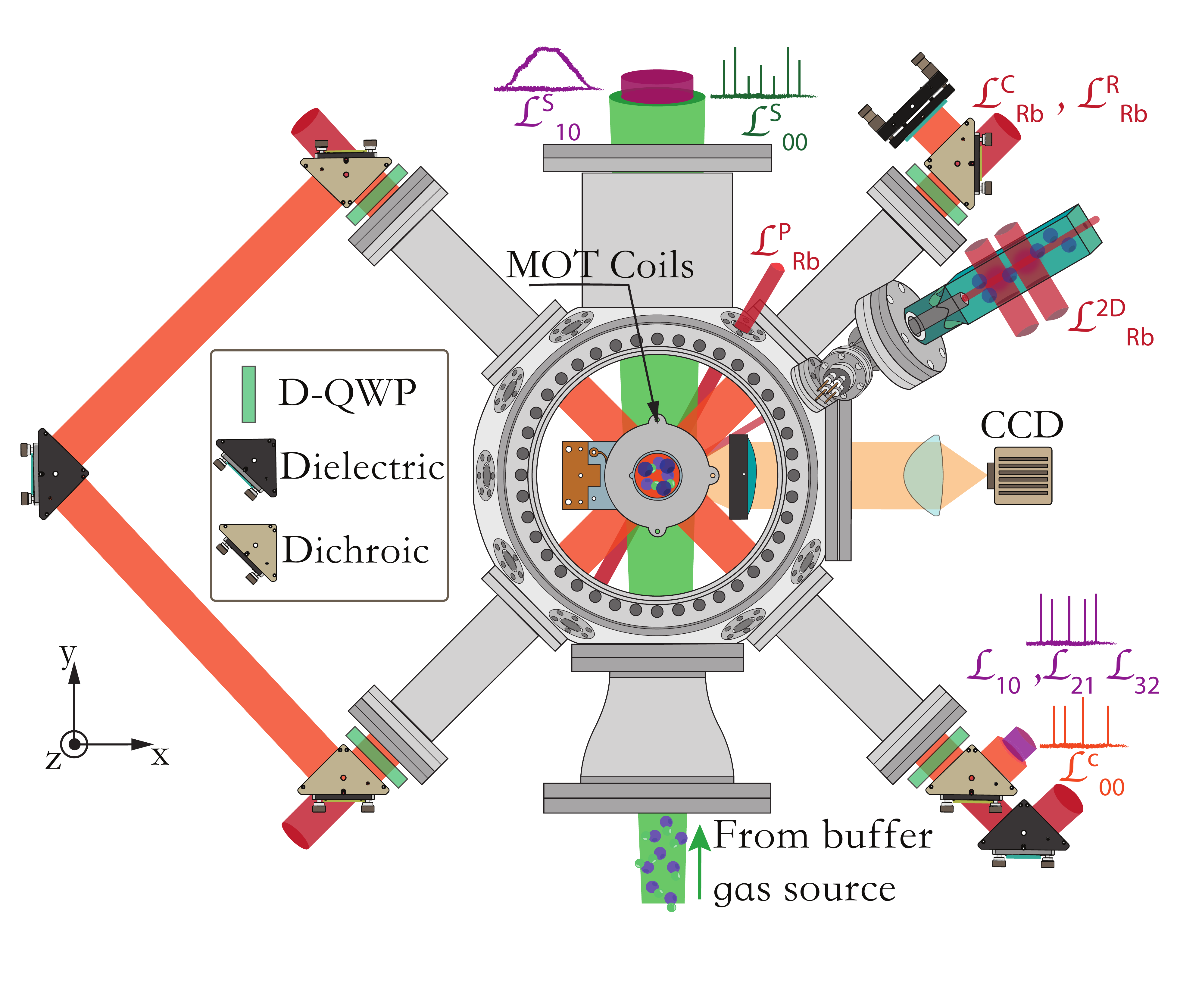}
    \caption{Apparatus for creating the dual-species MOT of CaF and Rb.  
    The light for the atom and molecule MOTs are combined using dichroic mirrors and propagate along the same path through the vacuum chamber. Dichroic quarter wave plates (D-QWP) generate the required circular polarizations. The fluorescence from either species can be imaged onto a CCD camera. The atom cloud can also be measured by absorption imaging. The various beams are indicated using the following notation. $\mathcal{L}^{\rm C}_{\rm Rb}$: Rb 3D MOT cooling; $\mathcal{L}^{\rm R}_{\rm Rb}$: Rb 3D MOT repump; $\mathcal{L}^{\rm 2D}_{\rm Rb}$: Rb 2D MOT; $\mathcal{L}^{\rm P}_{\rm Rb}$: Rb absorption imaging probe beam; $\mathcal{L}^{\rm C}_{00}$: CaF MOT cooling; $\mathcal{L}_{10}$, $\mathcal{L}_{21}$, $\mathcal{L}_{32}$: CaF MOT repumps; $\mathcal{L}^{\rm S}_{00}$: CaF slowing; $\mathcal{L}^{\rm S}_{10}$: CaF slowing repump. Also indicated are the frequency spectra of the lasers for CaF.
    \label{fig:setup}}
\end{figure}

Figure \ref{fig:setup} illustrates the setup we use to create the dual-species MOT of Rb and CaF. Coils inside the vacuum chamber create a quadrupole magnetic field with an axial gradient of 30~G~cm$^{-1}$. A 3D MOT of $^{87}$Rb atoms is formed using  six independent laser beams. Each beam has a $1/e^{2}$ radius of 1~cm and contains around 20~mW of cooling light red-detuned from the $F=2\rightarrow F'=3$ transition, and 3~mW of repump light resonant with the $F=1\rightarrow F'=2$ transition. The 3D MOT is loaded from a slow, collimated atomic beam extracted from a 2D MOT, which is formed in a glass cell attached to the main chamber. The scheme used to make the CaF MOT is the same as described in previous work, so is summarized only briefly here. A pulsed beam of CaF is produced by a cryogenic buffer gas source~\cite{Truppe2017c}, slowed to low velocity using a frequency-chirped counter-propagating laser beam~\cite{Truppe2017}, and then captured into the MOT~\cite{Truppe2017b, Williams2017}. The six CaF MOT beams are generated by passing the same beam six times through the chamber. Dichroic mirrors are used to combine and separate the Rb and CaF light, and dichroic waveplates are used to generate the required circular polarizations. 

Each experiment begins by loading the desired number of atoms into the 3D MOT at a typical load rate of $3\times 10^{9}$ atoms/s. We use between $9 \times 10^8$ and $8 \times 10^9$ atoms. Their temperature is 0.65~mK at the lower end of this range, rising to about 1~mK at the upper end. Once the atoms are loaded we switch off the 2D MOT and initiate the sequence to load the molecules, which takes 40~ms in total. We typically load about $10^{4}$ molecules. While the CaF MOT is being loaded, $I_{00}$ is set to its maximum value of $I_{00}^{\rm max}=1000$~mW~cm$^{-2}$.  Then, unless stated otherwise, $I_{00}$ is ramped down to $0.2 I_{00}^{\rm max}$; this lowers the photon scattering rate, reduces the temperature of the molecules from 10~mK to 3.5~mK, and increases the MOT lifetime. At this intensity, the trap oscillation frequency is about 90~Hz and the trap depth is approximately 75~mK~\cite{Williams2017}.

With both species trapped we acquire a set of 32 fluorescence images of the molecule cloud using a CCD camera. The images are separated in time by 18~ms, each has an exposure time of 10~ms, and the start of the first image defines $t=0$. We also record a set of background images where everything is identical except that no molecules are loaded. After subtracting these background images from the signal images, we determine the number of molecules remaining as a function of time, $t$, and thus the loss rate from the MOT in the presence of the atoms, $\Gamma_2$. Separately, we measure the loss rate of molecules in the absence of atoms, $\Gamma_{1}$. In this case, the experimental sequence is identical except that the Rb repump light is turned off for both the 2D and 3D MOTs, so that no atoms are loaded. $\Gamma_{1}$ is determined by the rate of decay to states that lie outside the laser cooling scheme, and the rate of evaporation from the high energy tail of the distribution~\cite{Williams2017}. The difference between $\Gamma_1$ and $\Gamma_2$ is the atom-induced loss rate $\Gamma_{\rm Rb-CaF}$.

\section{Results}

\begin{figure}[t!]
\centering
    \includegraphics[width=0.7\textwidth]{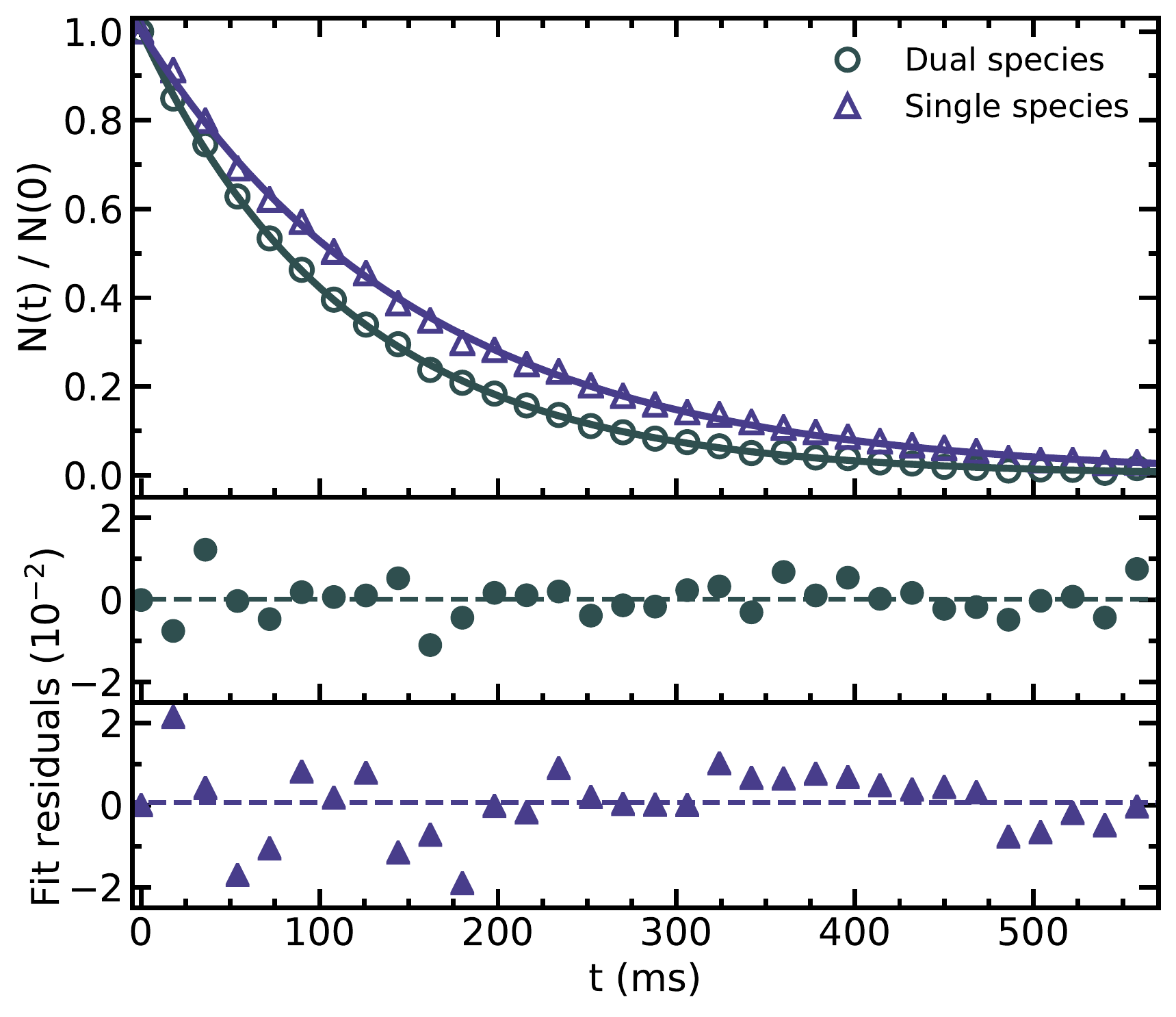}
    \caption{Fraction of molecules remaining in the MOT as a function of time, both with (green circles) and without (blue triangles) atoms. The lines are fits to $N(t)/N(0) = e^{-\Gamma t}$, and determine $\Gamma_1$ and $\Gamma_2$. The lower two panels show the fit residuals and the dashed lines indicate their averages.
    \label{fig:lifetime}}
\end{figure}

Figure \ref{fig:lifetime} shows the number of molecules as a function of time, normalized to the number at $t=0$, both with and without atoms. We fit these data to a single exponential decay model with the rate as the only free parameter. For the data shown, where the density of atoms is $(5.2 \pm 1.0) \times 10^{10}$~cm$^{-3}$, we find $\Gamma_{1}= 6.39$~s$^{-1}$ and $\Gamma_{2} = 8.6$~s$^{-1}$. For each experiment, these loss rates are measured 15 times, and this set of values is used to determine the mean value of $\Gamma_{\rm Rb-CaF}$, together with its standard error. The lifetime of the Rb MOT is around 8~s, which is at least 50 times longer than that of the CaF MOT, so the decay in the number of atoms during the measurement has a negligible effect on the loss rate measurements. In addition, the number density of atoms exceeds that of the molecules by 6 orders of magnitude, so loss of atoms due to collisions with molecules may safely be ignored.

The collision-induced loss rate is related to the inelastic rate coefficient, $k_2$, by
\begin{equation}
  \Gamma_{\rm Rb-CaF} = k_2 N_{\rm Rb} {\cal F}  
\end{equation}
where $N_{\rm Rb}$ is the number of atoms and
\begin{equation}
    {\cal F} = \int f_{\rm Rb}(\vec{r})f_{\rm CaF}(\vec{r})d^3\vec{r}
\end{equation}
is the overlap integral between the density distributions of the two species, $f_{s}(\vec{r})$, which are normalized such that $\int f_{s}(\vec{r})d^3\vec{r} = 1$. We measure $N_{\rm Rb}$ by absorption imaging, as discussed in \ref{sec:absorption_measurement}. We measure the density distributions along $y$ and $z$ by imaging the fluorescence of both clouds onto the same CCD camera. We select the fluorescence from CaF using a band-pass filter centred near 606~nm. The fluorescence from Rb is much brighter, so we attenuate it using a neutral density filter. We find that the centres of the two clouds are displaced by 1.5~mm along $y$, but coincide along $z$. We attribute this displacement to imperfect polarization of all the beams, and an intensity imbalance between the counter-propagating CaF MOT beams. These imperfections are consequences of the optical setup (see Fig.~\ref{fig:setup}), and are not easily corrected.

\begin{figure}[t!]
\centering
    \includegraphics[width=0.7\textwidth]{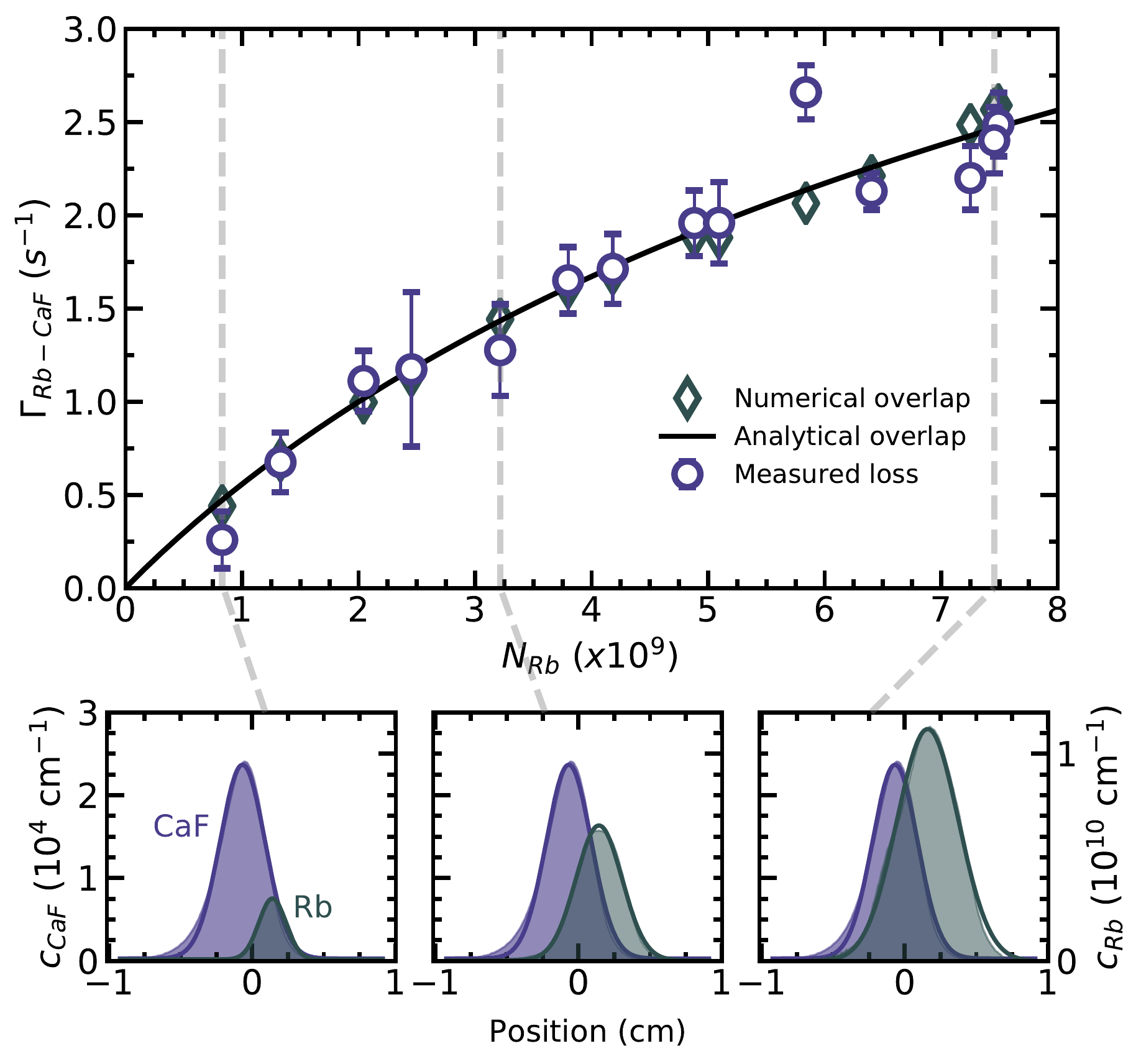}
    \caption{Determining the loss rate coefficient. The main figure shows $\Gamma_{\rm Rb-CaF}$ as a function of $N_{\rm Rb}$. Circles: measurements. Line: Best fit to the analytical model described in the text. Diamonds: best fit to the numerical model described in the text. The lower three panels show the density distributions integrated over $x$ and $z$ ($c_{\rm CaF}$ and $c_{\rm Rb}$), showing how the overlap between the two clouds in the $y$ direction changes with $N_{\rm Rb}$.
    \label{fig:loss_rate_coefficient}}
\end{figure}

The circular points in Fig.~\ref{fig:loss_rate_coefficient} show measurements of $\Gamma_{\rm Rb-CaF}$ as a function of $N_{\rm Rb}$. The data do not lie on a straight line because the overlap integral changes with $N_{\rm Rb}$. This can be seen in the lower panels of the figure, which show the density distributions along $y$ for three different values of $N_{\rm Rb}$. These are measured for every $N_{\rm Rb}$, so the overlap integrals along $y$ and $z$ are known. We cannot measure the distributions along $x$ since there is no optical access in this direction. Instead, we make the reasonable assumption that the MOTs are radially symmetric. This leaves a single undetermined parameter, $\Delta x$, the displacement between the two clouds along $x$. Fitting to the data in Fig.~\ref{fig:loss_rate_coefficient} determines both $k_2$ and $\Delta x$, as we now explain.

We fit two models to the data in Fig.~\ref{fig:loss_rate_coefficient}. They differ in the method used to determine ${\cal F}$. In the first method, which we call the analytical model, the clouds are described by Gaussian distributions in all three directions. In this case, 
\begin{equation}
    \mathcal{F} = \frac{1}{(2\pi)^{3/2}} \displaystyle\prod_{i\in \{x,y,z\}} \frac{1}{\tilde{\sigma}_i}\exp\left(\frac{-\Delta i^{2}}{2\tilde{\sigma}_i^2}\right)
    \label{eq:analytical_overlap_integral}
\end{equation}
where $\tilde{\sigma}_i^2 = (\sigma_{i}^{\rm Rb})^{2} + (\sigma_{i}^{\rm CaF})^{2}$. Here, $\sigma^{\rm Rb}_i$ and $\sigma^{\rm CaF}_i$ are the rms widths and $\Delta i$ are the displacements.
Fitting Gaussians to the measured distributions yields all these parameters at each value of $N_{\rm Rb}$, except $\Delta x$. The values of $\sigma^{\rm CaF}_{y,z}$ show no dependence on $N_{\rm Rb}$. Neither do $\Delta y$ and $\Delta z$, and we assume that the same is true for $\Delta x$. It is convenient (though not necessary) to express $\sigma^{\rm Rb}_{y,z}$ as continuous functions of $N_{\rm Rb}$, and we find that a simple power-law model, $\sigma^{\rm Rb}_{i} = \alpha_{i} N_{\rm Rb}^{\gamma_{i}}$ describes the data well. In this way, we obtain ${\cal F}(N_{\rm Rb})$ with $\Delta x$ as the only free parameter. Fitting this model to the data in Fig.~\ref{fig:loss_rate_coefficient} yields the line shown in the figure, with best-fit parameters $k_{2} = (1.29 \pm 0.21) \times 10^{-10}$~cm$^{3}$/s and $\Delta x = 0.13 \pm 0.07$~cm.

For large atom clouds, the distributions deviate from Gaussians. Our second model, which we call the numerical model, does a better job of accounting for this. We express the overlap integral as a product of two parts, ${\cal F} = \Upsilon^{\rm 2D} {\cal G}$. 
Here, $\Upsilon^{\rm 2D}$ is the overlap integral in 2D evaluated by multiplying the two images pixel by pixel,
\begin{equation}
    \Upsilon^{\rm 2D} = \frac{1}{\Delta A} \sum_{j} \mathcal{P}_{j}^{\rm Rb} \cdot \mathcal{P}_{j}^{\rm CaF}.
    \label{eq:numerical_overlap_integral}
\end{equation}
The sum is over all pixels, $\Delta A$ is the area in the object plane represented by one pixel, which is determined experimentally, and
\begin{equation}
    \mathcal{P}_{j}^{s} = \frac{p_{j}^{s}}{\displaystyle\sum_{k} p_{k}^{s}}
    \label{eq:normalized_pixels}
\end{equation}
where $p_j^{s}$ is the signal of species $s$ in pixel $j$. In the $x$ direction, we assume Gaussian distributions as before; $\mathcal{G}$ is their overlap integral. This model is defined only at the values of $N_{\rm Rb}$ used in the measurement. The diamonds in Fig.~\ref{fig:loss_rate_coefficient} are the results of fitting this model to the data. The best-fit parameters are $k_{2} = (1.43 \pm 0.29) \times 10^{-10}$~cm$^{3}$/s and $\Delta x = 0.13 \pm 0.07$~cm. We take these results to be the more accurate ones, but note that they are entirely consistent with the ones found from the analytical model, showing that it makes little difference whether we use the actual distributions or Gaussian fits to those distributions. 

\begin{figure}[t]
\centering
    \includegraphics[width=1.0\textwidth]{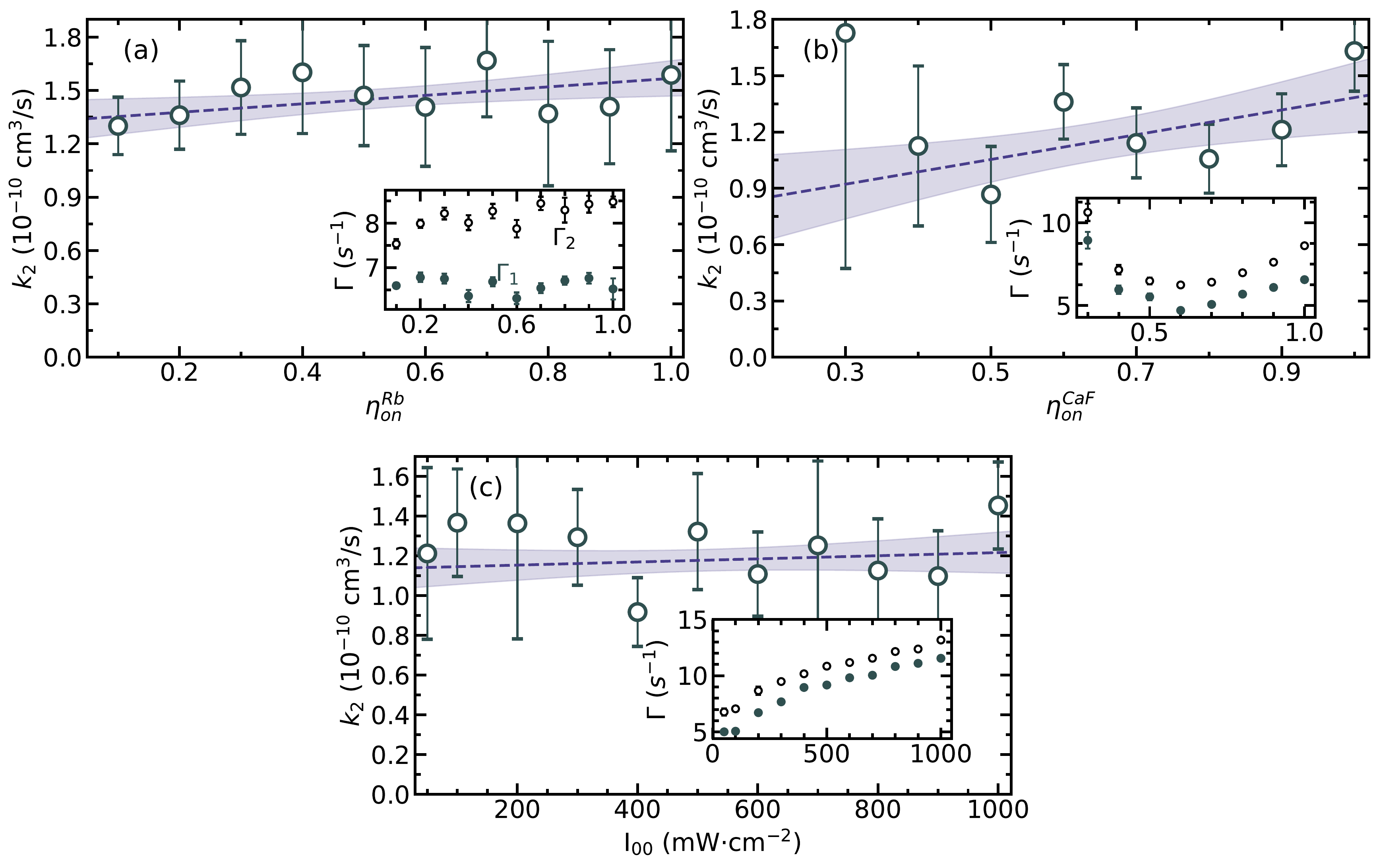}
    \caption{Effect of MOT light on loss rate coefficient. (a) Rb MOT light (780~nm) is modulated at 10~kHz and $k_2$ measured as a function of the duty cycle, $\eta^{\rm Rb}_{\rm on}$. (b) CaF MOT light (606~nm) is modulated at 10~kHz and $k_2$ measured as a function of the duty cycle, $\eta^{\rm CaF}_{\rm on}$. (c) $k_2$ as a function of the intensity of the CaF MOT light, $I_{00}$. Dashed lines are linear fits and shaded regions are 95\% confidence bands. Error bars are $1\sigma$ statistical uncertainties. Insets show how $\Gamma_1$ (filled circles) and $\Gamma_2$ (open circles) vary in each experiment. For all the experiments, the number of atoms is $N_{\rm Rb} = 3.5 \times 10^9$.
    \label{fig:light_experiments}}
\end{figure}

Next, we turn to the question of the process responsible for the observed collisional loss. The dual-species MOT is a mixture of atoms in the ground ($^{2}{\rm S}$) and excited ($^{2}{\rm P}_{3/2}$) states with molecules in the ground ($^{2}\Sigma$) and excited ($^{2}\Pi_{1/2}$) states. The dominant inelastic process could involve any combination of these. There can also be light-induced processes due either to the light near-resonant with the atoms, or the light near-resonant with the molecules. Several hyperfine states are involved too. By measuring $k_2$ under various conditions, we disentangle the possible mechanisms. We independently modulate the intensities of the Rb and CaF MOT light, using a fixed modulation amplitude of 100\%, a fixed frequency of 10~kHz, and a variable duty cycle, $\eta_{\rm on}^s$, where $s$ denotes the species. The modulation frequency is much faster than the trap oscillation frequencies (0.1--1~kHz) but much slower than the rate at which the state populations adjust to a change of intensity ($>1$~MHz). The effective restoring force in the trap, the trap depth, the time-averaged excited-state fraction, and any light-induced inelastic rates, must all be proportional to $\eta_{\rm on}^s$.

Figure \ref{fig:light_experiments}(a) shows how $k_2$ varies with the duty cycle of the Rb trap light, $\eta^{\rm Rb}_{\rm on}$. We use the method described above to determine $k_2$ in each case. The size of the Rb cloud changes with $\eta^{\rm Rb}_{\rm on}$. For example, the radial and axial rms radii of the cloud change from $\{\sigma_y,\sigma_z\} = \{1.78, 1.66\}$~mm when $\eta^{\rm Rb}_{\rm on}=1$ to $\{2.38, 2.2\}$~mm when $\eta^{\rm Rb}_{\rm on} = 0.1$, with the largest change for duty cycles between 0.3 and 0.1. This change reduces the overlap between the clouds at low duty cycles, which in turn decreases $\Gamma_2$, as can be seen in the inset of Fig.~\ref{fig:light_experiments}(a). The position of the cloud varies by less than 150~$\mu$m, while the number of atoms remains constant. Using the measured density distributions, we determine the overlap integral, ${\cal F}$, for each value of $\eta^{\rm Rb}_{\rm on}$, using the analytical model. From the values of $\Gamma_1$, $\Gamma_2$, $N_{\rm Rb}$ and ${\cal F}$, we obtain the values of $k_2$ shown in the main part of the figure. Fitting these data to a linear model gives a gradient of $(2.4 \pm 3.0) \times 10^{-11}$~cm$^{3}$/s, which is consistent with zero. The hypothesis that the observed loss is due entirely to collisions with excited-state atoms can be rejected with very high confidence since the intercept of the linear fit is $9.5\sigma$ away from zero.
The hypothesis that the loss is due entirely to processes induced by the Rb cooling light can be rejected at the same level. If both ground-state and excited-state atoms are involved, we can write 
\begin{equation}
    k_2 = k_2^{\rm e} f_{\rm e} + k_2^{\rm g}(1-f_{\rm e}) = k_2^{\rm g} + f_{\rm e}^{\rm max} \eta^{\rm Rb}_{\rm on}(k_2^{\rm e} - k_2^{\rm g}),
    \label{eq:ke_kg}
\end{equation}
where $k_2^{\rm g}$ and $k_2^{\rm e}$ are the loss rate coefficients for collisions with ground-state and excited-state Rb atoms, and $f_{\rm e}$ is the excited-state fraction whose value at $\eta^{\rm Rb}_{\rm on} = 1$ is $f_{\rm e}^{\rm max}$. The excited-state fraction in a Rb MOT has been studied experimentally over a wide range of parameters in Ref.~\cite{Shah2007}. For our parameters (detuning of $-3\Gamma$, intensity of $72.5$~mW~cm$^{-2}$), the result is $f_{\rm e}^{\rm max} \simeq 0.15$. From the linear fit, we obtain $k_2^{\rm g} = (1.33 \pm 0.14) \times 10^{-10}$~cm$^{3}$~s$^{-1}$ and $k_2^{\rm e} = (2.9 \pm 1.9) \times 10^{-10}$~cm$^{3}$~s$^{-1}$. 

Figure \ref{fig:light_experiments}(b) shows how $\Gamma_1$, $\Gamma_2$ and $k_2$ vary as we change the duty cycle of the CaF MOT light, $\eta^{\rm CaF}_{\rm on}$. The Rb MOT light is on continuously. In these experiments, the rms radii of the CaF cloud change from $\{\sigma_y,\sigma_z\} = \{1.40, 1.05\}$~mm at $\eta^{\rm CaF}_{\rm on}=1$ to $\{1.96, 1.28\}$~mm at $\eta^{\rm CaF}_{\rm on}=0.3$. The change in the cloud position is less than $0.2$~mm and has no significant effect.  As the duty cycle is lowered towards 0.6, both $\Gamma_1$ and $\Gamma_2$ decrease. This is due to the reduction in the time-averaged scattering rate. For duty cycles below 0.6 the loss rate increases again because the trap depth becomes too small to confine the higher-energy molecules. The linear fit to the $k_2$ data has a gradient of $(7 \pm 5) \times 10^{-11}$~cm$^{3}$/s, which is consistent with zero. The intercept is $2\sigma$ away from zero, so the hypothesis that the observed loss is due entirely to collisions with excited-state molecules seems unlikely.

In addition to changing the duty cycles, we measure the loss rate coefficient over a wide range of CaF MOT light intensities, $I_{00}$. Figure \ref{fig:light_experiments}(c) shows $\Gamma_1$, $\Gamma_2$ and $k_2$ for values of $I_{00}$ between $I^{\rm max}_{00} = 1000$~mW/cm$^{2}$ and $0.05 I^{\rm max}_{00}$. From the MOT fluorescence, we find that the excited-state population changes by a factor of 5 over this range.  Lowering the intensity over this range reduces the geometric mean size of the cloud by 14\% due to a decrease in temperature, and reduces the displacement between the two clouds by 0.6~mm. 
Fitting the data to a linear model, we find a gradient of $(0.8 \pm 2.7) \times 10^{-14}$~cm$^{5}$/(s~mW), showing that the trap light has no effect on the loss at this level of uncertainty. The intercept is $6.7\sigma$ from zero, so the hypothesis that the loss is due entirely to the presence of near-resonant light can be rejected with high confidence.

\section{Discussion}

In earlier work using CaF and Rb mixtures in a magnetic trap, a loss rate coefficient of $k_{2} = (6.6 \pm 1.5) \times 10^{-11}$~cm$^{3}$/s was measured for molecules in $N=1$ at a relative temperature of 133~$\mu$K~\cite{Jurgilas2021}. There was no significant dependence on the hyperfine state. This result was compared to the predictions of a single-channel model based on quantum defect theory~\cite{Frye2015}, which takes account of quantum reflection off both centrifugal barriers and the long-range attractive potential. The model is characterized by two parameters, a short-range phase shift $\delta^\mathrm{s}$ and a short-range loss parameter $y$, which it assumes are independent of energy and partial wave. The measured loss rate was close to the universal rate~\cite{Idziaszek2010}, where every collision that reaches short range results in loss ($y=1$). However, it was shown that such a rate can arise for values of $y$ as low as 0.03 for certain values of $\delta^\mathrm{s}$. 
The loss was attributed to rotational relaxation, which is driven by the anisotropy of the interaction potential. Such loss is expected to be fast (but not necessarily universal) when the anisotropy is large compared to the spacing between the rotational states. The kinetic energy released in a collision that changes $N$ from 1 to 0 is 20.5~GHz, which is large compared to the trap depth.

For the present measurements, the effective temperature of the relative motion is $T=\mu(T_{\rm Rb}/m_{\rm Rb}+T_{\rm CaF}/m_{\rm CaF})=2.4$~mK, where $T_a$ and $m_a$ are the temperature and mass of species $a$, and $\mu$ is their reduced mass.
At this temperature, the results of the single-channel model have almost no dependence on $\delta^\mathrm{s}$ because shape resonances are washed out by thermal and partial-wave averaging~\cite{Jachymski2014}. The loss rate can be approximated as 
\begin{equation}
    k_2(T) = P k_2^{\rm max}(T),
    \label{eq:nonuniv}
\end{equation}
where the probability $P$ of loss for collisions that reach short range is $P=4y/(1+y)^2$. Here $k_2^{\rm max}(T)$ is the thermally averaged universal loss rate, which depends only on the collisional reduced mass $\mu$ and the asymptotic interaction potential $-C_6R^{-6}$; using the value of $C_6$ estimated in Ref.\ \cite{Lim2015}, we calculate $k_2^{\rm max}=1.32 \times 10^{-10}$~cm$^{3}$/s at $T=2.4$~mK. 
Our experimental result is thus consistent with universal loss, as observed previously for rotation-changing collisions at lower temperature. The present experiment limits $P$ to values greater than 0.86, corresponding to $y>0.46$, and rules out the lower range of $y$ permitted by the experiments at 133 $\mu$K.

At the temperature of the present experiment, where 4 to 7 partial waves contribute, universal loss is well approximated by a Langevin capture model \cite{Frye2015}. In this model, all classical trajectories with sufficient energy to cross the centrifugal barrier are assumed to result in loss. For an asymptotic potential $-C_6R^{-6}$, the Langevin model gives \cite{Gorin1939,Fernandez-Ramos2006}
\begin{equation}
k_2^{\rm max}(T) = 2^{11/6}\, \Gamma(2/3) \left(\frac{\pi}{\mu}\right)^{1/2} C_6^{1/3} (k_\mathrm{B} T)^{1/6},
\end{equation}
where $\Gamma(z)$ is the gamma function. For Rb+CaF at $T=2.4$~mK, this evaluates to
$k_2^{\rm max}=1.34 \times 10^{-10}$~cm$^{3}$/s, in good agreement with both the experimental value and the quantum-mechanical universal loss rate. The Langevin rate scales as $T^{1/6}$; in the case of universal loss, this scaling remains valid down to temperatures around $T \sim E_6/k_\mathrm{B}$ \cite{Frye2015},\footnote{In the nonuniversal case, when $y\ll 1$, $T^{1/6}$ scaling can break down even at temperatures well above $E_6/k_\mathrm{B}$ \cite{Frye2015}.} where $E_6=\hbar^3/\sqrt{8\mu^3C_6}$, which is 124 $\mu$K for Rb+CaF. The ratio between the values of $k_2$ measured at 2.4~mK and 133~$\mu$K is consistent with this scaling.

It is interesting to compare our results to those found for atomic MOTs~\cite{Prentiss1988, Sesko1989, Wallace1992, Kawanaka1993, Marcassa1993, Ritchie1995, Gensemer1997}. Hyperfine-changing collisions between ground-state atoms are an important loss mechanism in atomic MOTs when the trap depth is smaller than the energy released in such collisions. Loss rate coefficients comparable to those measured here have been observed. As the trap depth increases, this mechanism is suppressed and the loss rate can be reduced by orders of magnitude. In our MOT, collisions that change only the hyperfine state of the molecule do not release enough energy for the molecule to be lost. However, in a low-energy collision that changes the hyperfine state of $^{87}$Rb, the molecule emerges with a velocity of 7.4~m/s. This is close to the capture velocity of the MOT at high intensity, which is about 11~m/s~\cite{Williams2018}. The trap depth decreases as the duty cycle is lowered, so we might have expected a strong dependence on the duty cycle, as loss due to hyperfine-changing collisions turns on. However, as argued above, rotational relaxation causes loss that is already at the universal limit.
In this limit, adding  an additional collisional loss mechanism, no matter how strong, does not increase the loss rate. The same argument applies to radiative escape, where a colliding pair absorbs a photon at intermediate range and emits a lower-energy photon at short range. 

Excited-state atoms and molecules form distinct populations with their own loss mechanisms and their own distinct value of the universal loss rate. In particular, there can be loss due to fine-structure-changing collisions with excited-state atoms, which is known to be important in atomic MOTs, and due to $\Lambda$-doublet-changing collisions involving excited-state molecules. The molecules are excited to the positive-parity component of a $\Lambda$ doublet, which lies 1.36~GHz higher in energy than the negative-parity component. A collision that changes this parity releases the energy and takes the molecule out of the laser-cooling cycle.
Such transitions are driven directly by odd-order terms in the Legendre expansion of the anisotropy of the interaction potential \cite{Klar1973}. These terms are large compared to the parity splitting, so are also expected to cause fast loss, which may be sufficient to reach the universal limit. The excited-state universal rate will differ from that for the ground state due to the different value of $C_6$. However, the difference is likely to be small because the Langevin rate depends only weakly on $C_6$, as $C_6^{1/3}$. The observed insensitivity of the loss rate to the CaF excited-state fraction is consistent with parity-changing collisions that are sufficient to cause universal loss.

\section{Conclusions}

We have developed a dual-species magneto-optical trap of CaF molecules and Rb atoms and measured the loss rate of CaF from the trap due to inelastic collisions with the atoms. At an effective collision temperature of 2.4~mK, the measured loss rate coefficient is $k_{2} = (1.43 \pm 0.29) \times 10^{-10}$~cm$^{3}$/s. This result is consistent with the universal loss rate and with the loss rate calculated in the classical limit using a Langevin capture model. The result constrains the probability of loss for collisions that reach short range to $P>0.86$. The corresponding constraint on the loss parameter is $y>0.46$. We expect $y$ to be independent of energy, so this constraint should apply also to collisions in the ultracold regime. Rotational relaxation of ground-state molecules colliding with ground-state atoms is sufficient to account for the loss rate observed. We cannot rule out contributions from excited-state atoms or molecules, or from light-induced processes, but these processes alone are insufficient to explain our observations.

Our dual-species MOT could be used as a starting point for forming the triatomic CaFRb molecule. The formation of diatomic molecules by photoassociation of atoms is often studied by adding an additional laser to a dual-species MOT and measuring resonances in the loss rate as the frequency of this laser is scanned. This is unlikely to work in our dual-species MOT since the loss rate is already close to the universal limit; adding a new loss mechanism cannot much increase the observed loss rate. However, the atoms and molecules could be loaded from the MOT into a conservative trap where molecule formation by photoassociation or magnetoassociation through a Feshbach resonance could be studied. 

Our results are also important for designing an experimental protocol for sympathetic cooling of molecules using evaporatively cooled atoms~\cite{Lim2015}. Again, the atoms and molecules first need to be loaded from the MOT into a conservative trap. The rapid collisional losses observed in the dual-species MOT will limit the density that can be achieved. This limitation could be avoided by loading molecules and atoms sequentially into the conservative trap. Collisional losses in a magnetic trap are much slower, provided that the molecules are in the rotational ground state~\cite{Jurgilas2021}. Laser cooling is now being applied to many molecular species (see \cite{Fitch2021} for a recent review) and the methods developed here could be used to produce a wide range of atom-molecule mixtures and to study how they interact at low temperatures.

\ack
We are grateful for expert technical assistance from Jon Dyne and David Pitman. We acknowledge helpful discussions with Micha{\l} Tomza. This work was supported by EPSRC grants EP/P01058X/1 and EP/V011499/1.

\appendix

\section{Measuring the number of atoms}
\label{sec:absorption_measurement}

We find the number of atoms present in the trap by acquiring absorption images of the cloud released from the MOT. The absorption probe is detuned from the $F=2 \rightarrow F'=3$ transition by $\Delta = -2.35 \Gamma$, where $\Gamma=3.81 \times 10^7$~rad~s$^{-1}$ is the linewidth of the transition. The light propagates in the direction of the magnetic field, whose magnitude is 400~mG. The probe laser is turned on for 100~$\mu$s and has an intensity of about $0.02 I_{\rm s}$, where $I_{\rm s}$ is the saturation intensity. The polarization of the probe is calibrated using a polarimeter. We choose a circular polarization so that only $\sigma^{+}$ transitions are driven. In this case, once the populations reach the steady-state distribution, the absorption cross section is $\sigma_{\rm ss} = (3\lambda^{2}/2\pi)(1+4\Delta^2/\Gamma^2)^{-1}$. The actual cross section differs from this value for two reasons as we explain below. 

\begin{figure}[t!]
\centering
    \includegraphics[width=0.6\textwidth]{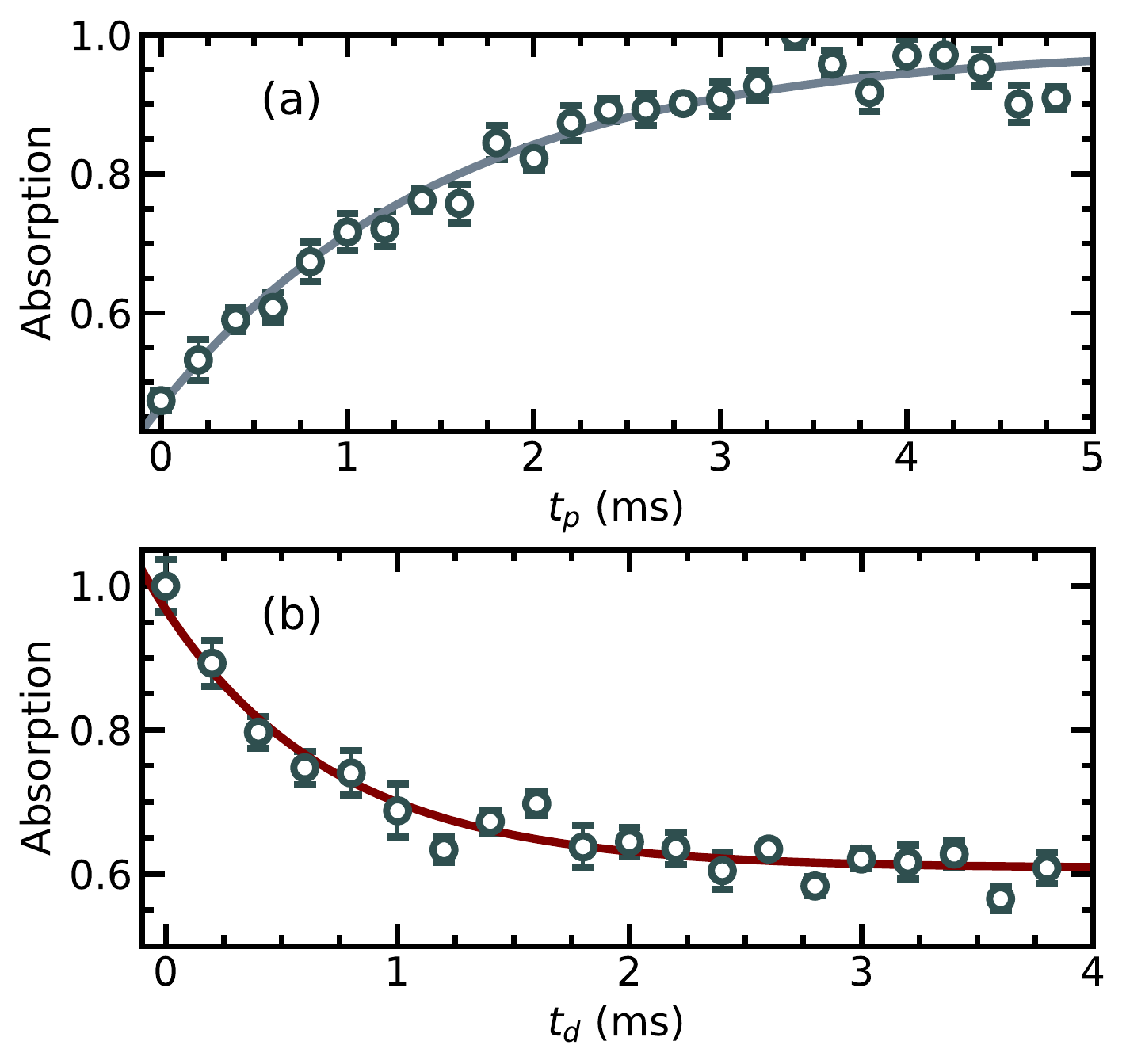}
    \caption{Absorption of the probe beam by the atom cloud released from the MOT. In (a) the probe beam is switched on 1~ms after turning off the MOT coils and kept on for a variable time t$_{\rm p}$ before an absorption image is acquired. In (b) the absorption is measured for different delay times t$_{\rm d}$ after turning off the coils. The solid lines in both figures are fits to exponential models.
    \label{fig:absorption_probe}}
\end{figure}

Atoms in the MOT are distributed across all the Zeeman sub-levels of $F=2$, whereas the absorption beam optically pumps them towards $m_F=2$, so it takes some time for the absorption cross section to reach its steady-state value. To measure the effect of this, we performed the following experiment. The probe beam is turned on 1~ms after switching off the MOT coils, and held on for a variable time $t_{\rm p}$ before triggering the camera to acquire the absorption image. Figure~\ref{fig:absorption_probe}(a) shows how the total absorption of the probe beam varies with $t_{\rm p}$. Fitting the data to an exponential model we find an optical pumping time constant of $1.54(13)$~ms. This is similar to the timescale found by solving rate equations describing the light-atom interaction. From the fit, we find that the absorption at $t_{\rm p}\rightarrow\infty$ is a factor of 2.10(4) higher than at $t_{p}=0$. This result is consistent with an initially uniform distribution of atoms over the 5 ground-state sub-levels of $F=2$, since the average absorption cross section in that case is $\frac{7}{15}\sigma_{\rm ss}$. In the collision experiments, the absorption image is acquired with t$_{\rm p}=0$, so we use an absorption cross section of $\sigma_{\rm ss}/2.1$ when we estimate the number of atoms.

The second effect that can change the absorption cross section is the changing magnetic field in the experiment, since the absorption image is acquired only 1~ms after turning off the MOT coils. Figure~\ref{fig:absorption_probe}(b) shows how the total absorption of the probe beam varies with the delay time t$_{\rm d}$ between switching off the magnetic field and acquiring the absorption image. The absorption is highest at $t_{\rm d}=0$, presumably because the MOT magnetic field tends to shift the atoms closer to resonance with the light. Fitting an exponential model to these data gives a time constant of $0.73(9)$~ms. In the collision experiments, we acquire images at $t_{\rm d}=1$~ms so as to avoid excessive expansion of the atom cloud. The absorption is about 19\% higher at this time compared to long times when the magnetic field has decayed away and we account for this when we estimate the number of atoms.

\section*{References}
\bibliographystyle{iopart-num}
\bibliography{references}

\end{document}